# Evaluating the Impact of Using GRASP Framework on Clinicians and Healthcare Professionals' Decisions in Selecting Clinical Predictive Tools

Mohamed Khalifa [1], Farah Magrabi [1] and Blanca Gallego [2]

[1] Australian Institute of Health Innovation, Faculty of Medicine and Health Sciences, Macquarie University, Sydney, Australia

[2] Centre for Big Data Research in Health, Faculty of Medicine, University of New South Wales, Sydney, Australia

**Authors:**

Mohamed Khalifa: mohamed.khalifa@mq.edu.au

Farah Magrabi: farah.magrabi@mq.edu.au

Blanca Gallego: b.gallego@unsw.edu.au

**Abstract**

**Background:** When selecting predictive tools, clinicians and healthcare professionals are challenged with an overwhelming number of tools, most of which have never been evaluated for comparative effectiveness. To overcome this challenge, the authors developed and validated an evidence-based framework for grading and assessment of predictive tools (GRASP), based on the critical appraisal of published evidence.

**Methods:** To examine GRASP impact on professionals' decisions, a controlled experiment was conducted through an online survey. Randomising two groups of tools and two scenarios; participants were asked to select the best tools; most validated or implemented, with and without GRASP. A wide group of international participants were invited. Task completion time, rate of correct decisions, rate of objective vs subjective decisions, and level of decisional conflict were measured.

**Results:** Valid responses received were 194. Compared to not using the framework, GRASP significantly increased correct decisions by 64% (T=8.53, p<0.001), increased objective decision making by 32% (T=9.24, p<0.001), and decreased subjective decision making; based on guessing and based on prior knowledge or experience by 20% (T=-5.47, p<0.001) and 8% (T=-2.99, p=0.003) respectively. GRASP significantly decreased decisional conflict; increasing confidence and satisfaction of participants with their decisions by 11% (T=4.27, p<0.001) and 13% (T=4.89, p<0.001) respectively. GRASP decreased task completion time by 52% (T=-0.87, p=0.384). The average system usability scale of GRASP was very good; 72.5%, and 88% of participants found GRASP useful.

**Discussion and Conclusions:** Using GRASP has positively supported and significantly improved evidence-based decision making and increased accuracy and efficiency of selecting predictive tools. GRASP represents a high-level approach and an effective, evidence-based, and comprehensive, yet simple and feasible, method to evaluate, compare, and select clinical predictive tools.

**Keywords:** Clinical Prediction, Clinical Decision Support, Grading and Assessment, Evidence-Based, Impact Evaluation.

## 1. Background

Clinical decision support (CDS) systems have been proved to enhance evidence-based practice and support cost-effectiveness [1-6]. Based on Shortliffe's three levels classification, clinical predictive tools, here referred to simply as predictive tools, belong to the highest CDS level; providing patient-specific recommendations based on clinical scenarios, which usually follow clinical rules and algorithms, cost benefit analysis, or clinical pathways [7, 8]. Such tools include various applications; ranging from the simplest manual clinical prediction rules to the most sophisticated machine learning algorithms [9, 10]. These research-based applications provide diagnostic, prognostic, or therapeutic decision support. They quantify the contributions of relevant patient characteristics to derive the likelihood of diseases, predict their courses and possible outcomes, or support the decision making on their management [11, 12].

When selecting predictive tools for implementation at the clinical practice or for recommendation in clinical guidelines; clinicians and healthcare professionals, here referred to simply as "Professionals", involved in the decision making, are challenged with an overwhelming and ever-growing number of tools. Many of these have never been implemented or evaluated for comparative effectiveness [13-15]. By definition, healthcare professionals include all clinicians who provide direct care to patients, in addition to professionals who work in laboratories, researchers, and public health experts [16]. Professionals usually rely on previous experience, subjective evaluation or recent exposure to predictive tools in making selection decisions. Objective methods and evidence-based approaches are rarely used in such decisions [17, 18].

When developing clinical guidelines, some professionals search the literature for studies that describe development, implementation or evaluation of predictive tools. Others look for systematic reviews comparing tools' performance or development methods. However, there are no available approaches to objectively summarise or interpret such evidence [19, 20]. In addition, predictive tools selection decisions are time consuming; seeking a consensus of subjective expert views [21]. Furthermore, when experts make their decisions subjectively they face much decisional conflict; being less confident in the decisions they make and sometimes less satisfied with them [22].

To overcome this major challenge, the authors have developed and validated a new evidence-based framework for grading and assessment of predictive tools (The GRASP Framework) [23]. The aim of this framework is to provide standardised objective information on predictive tools to support the search for and selection of effective tools. Based on the critical appraisal of published evidence, GRASP uses three dimensions to grade predictive tools: 1) Phase of Evaluation, 2) Level of Evidence and 3) Direction of Evidence.

**Phase of Evaluation:** Assigns A, B, or C based on the highest phase of evaluation. If a tool's predictive performance, as reported in the literature, has been tested for validity, it is assigned phase C. If a tool's usability and/or potential effect have been tested, it is assigned phase B. Finally, if a tool has been implemented in the clinical practice, and there is published evidence evaluating its post-implementation impact, it is assigned phase A.

**Level of Evidence:** A numerical score, within each phase, is assigned based on the level of evidence associated with each tool. A tool is assigned grade C1 if it has been tested for external validity multiple times, grade C2 if it has been tested for external validity only once, and grade C3 if it has been tested only for internal validity. Grade C0 means that the tool did not show sufficient internal validity to be used in the clinical practice. Grade B1 is assigned to a predictive tool that has been evaluated, during the planning for implementation, for both of its potential effect, on clinical effectiveness, patient safety or healthcare efficiency, and for its usability. Grade B2 is assigned to a predictive tool that has been evaluated only for its potential effect, while if it has been studied only for its usability, it is assigned grade B3. Finally, if a predictive tool had been implemented then evaluated for its post-implementation impact, on clinical effectiveness, patient safety or healthcare efficiency, then it is assigned grade A1 if there is at least one experimental study of good quality evaluating its post-implementation impact, grade A2 if there are observational studies evaluating its impact, and grade A3 if the post-implementation impact has been evaluated only through subjective studies, such as expert panel reports.

**Direction of Evidence:** For each phase and level of evidence, a direction of evidence is assigned based on the collective conclusions reported in the studies. The evidence is considered positive if all studies about a predictive tool reported positive conclusions and negative if all studies reported negative or equivocal conclusions. The evidence is considered mixed if some studies reported positive and some reported either negative or equivocal conclusions. To decide an overall direction of evidence, a protocol is used to sort the mixed evidence into supporting an overall positive or negative conclusion. The protocol is based on two main criteria; 1) Degree of matching between the evaluation study conditions and the original tool specifications, and 2) Quality of the evaluation study. Studies evaluating tools in closely matching conditions to the tools' specifications and providing high quality evidence are considered first for their conclusions in deciding the overall direction of evidence.

The final grade assigned to a tool is based on the highest phase of evaluation, supported by the highest level of positive evidence, or mixed evidence that supports a positive conclusion. The GRASP framework concept is shown in Figure 1 and the GRASP framework detailed report is presented in Table 3 in the Appendix.

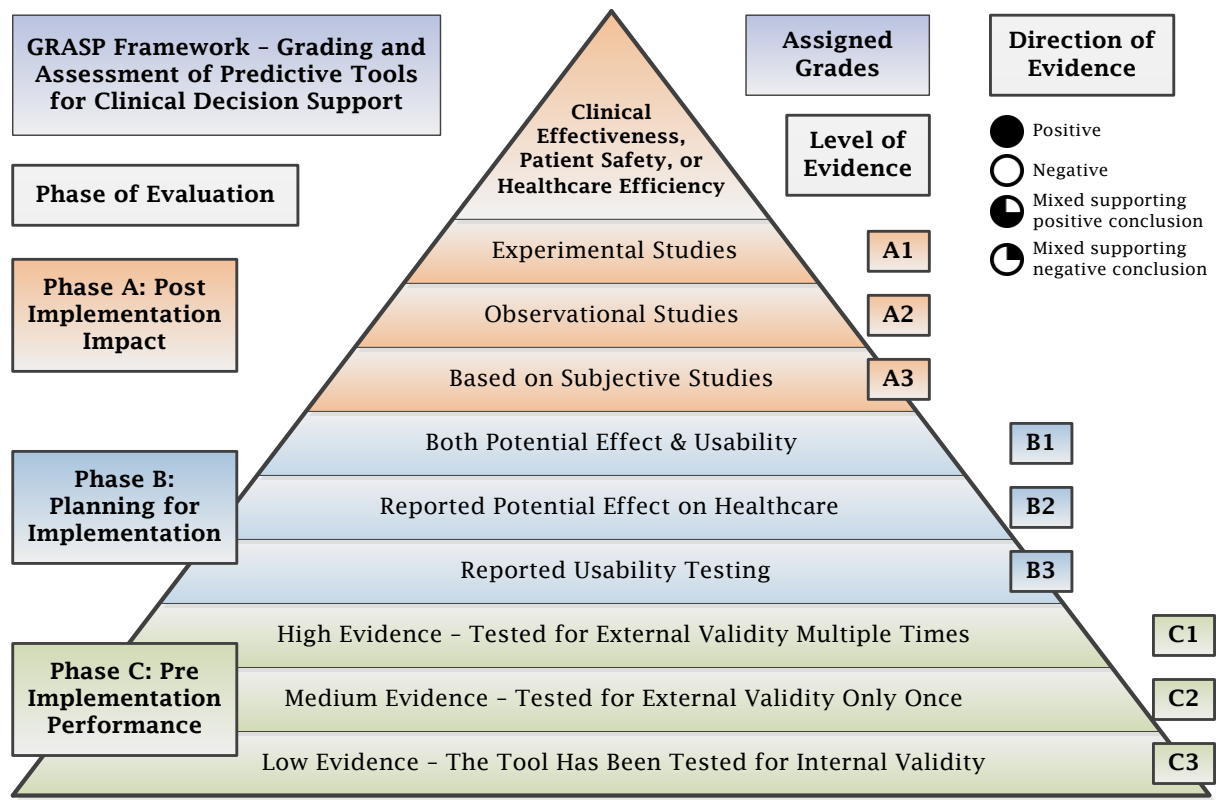

Figure 1: The GRASP Framework Concept [23]

The aim of this study is to evaluate the impact of using GRASP on the decisions made by professionals in selecting predictive tools for clinical decision support. The objective is to explore whether the GRASP framework is going to positively support professionals' evidence-based decision-making and improve their accuracy and efficiency in selecting clinical predictive tools. To explore this impact, a group of hypotheses have been proposed including that; using the GRASP framework by professionals is going to 1) Make their decisions more accurate, i.e. selecting the best predictive tools. 2) Make their decisions more objective, informed, and evidence-based, i.e. decisions are based on the information provided by the framework. 3) Make their decisions less subjective, i.e. decisions are less based on guessing, prior knowledge, or experience. 4) Make their decisions more efficient, i.e. decisions are made in less time. 5) Make them face less decisional conflict, i.e. become more confident in their decisions and more satisfied with them. We also propose that using GRASP can move professionals who have less knowledge, less experience, and are less familiar with predictive tools to an equal or even higher accuracy of decision making than professionals who have more knowledge, more experience, and are more familiar with tools, when they do not use GRASP.

## 2. Methods

### *2.1. The Study Design*

This study is based on experimental methods. It aims at examining the performance and outcomes of professionals' decisions in selecting predictive tools with and without using the GRASP framework. Through an online survey, the experiment involves asking participants to select the best predictive tool, for implementation at the clinical practice or for recommendation in clinical practice guidelines, from a group of five similar tools doing the same predictive task, one time with and another time without using the GRASP framework. In addition, participants are asked a few questions regarding the process of making their decisions through the two scenarios. Participants are also requested to provide their feedback on the perceived usability and usefulness of the evidence-based summary of the GRASP framework.

The emergency department (ED) is among the top healthcare specialties that are increasingly utilising predictive tools especially in the area of managing traumatic brain injury (TBI), being the leading cause of death and disability among trauma patients [24-27]. Two groups of predictive tools designed to exclude TBI in ED were prepared. The first group includes five tools for predicting TBI in paediatrics; PECARN – Paediatric Emergency Care Applied Research Network head injury rule, CHALICE – Children's Head injury ALgorithm for the prediction of Important Clinical Events, CATCH – Canadian Assessment of Tomography for Childhood Head injury rule, Palchak head injury rule, and Atabaki head injury rule [28-32].

The PECARN is the best tool among the five; being the most validated and the only tool that has been implemented in clinical practice and proved to have positive post-implementation impact [33, 34]. The second group includes five tools for predicting TBI in adults; CCHR – Canadian CT Head Rule, NOC – New Orleans Criteria, Miller criteria for head computed tomography, KHR – Kimberley Hospital Rule, and Ibanez model for head computed tomography [35-39]. The CCHR and NOC are the best tools among the five; being the only tools that have been implemented in clinical practice and are the most validated, showing high predictive performance [40-42].

Two scenarios were prepared for this experiment. The first is the control scenario, which includes providing participants with the basic information about each tool, the full text of the original studies describing the tools, in addition to allowing them to search the internet for information. The second is the experiment scenario, which includes providing participants with the main component of the GRASP framework, which is the evidence-based summary on the predictive tools, the full GRASP report on each tool, in addition to allowing them to search the internet for information.

To minimise bias, eliminate pre-exposure effect, and improve the robustness, the experiment includes randomising the two groups of predictive tools and the two scenarios. Accordingly, participants go randomly through one of four scenarios; 1) Paediatric Tools without GRASP then Adult Tools with GRASP, 2) Paediatric Tools with GRASP then Adult Tools without GRASP, 3) Adult Tools without GRASP then Paediatric Tools with GRASP, 4) Adult Tools with GRASP then Paediatric Tools without GRASP. Figure 2 shows the survey workflow and the randomisation.

The authors recruited a wide group of international professionals to participate in this experiment through an online survey. To identify potential participants who work at the emergency department and those who have knowledge or experience about CDS tools, published studies were used to retrieve the authors' emails and invite them. Four databases were used; MEDLINE, EMBASE, CINAHL, and Google Scholar to retrieve studies on CDS systems, tools, models, algorithms, pathways, or rules used in the emergency department, emergency service, or emergency medicine published over the last five years by professionals who work in the emergency departments or services of their healthcare organisations or those who conducted emergency medicine, emergency department, or emergency services research.

The authors expected the response rate to be around 10%. Before the deployment of the survey a pilot testing, by ten expert professionals, was conducted. The feedback of the pilot testing was used to improve the survey. Professionals who participated in the pilot testing were excluded from the participation in the final survey. An invitation email, introducing details about the study objectives, the GRASP framework, the experiment task, the survey completion time, which was estimated at 20 minutes, and a participation consent, was submitted to the identified potential participants with the link to the online survey. A reminder email, in two weeks, was sent to the potential participants who did not respond or complete the survey.

*2.2. The Study Survey*

The online survey was developed using Qualtrics Experience Management Solutions Platform [43]. The survey, illustrated through screenshots in the Appendix, includes five sections. The first section includes an introduction about the study objectives, the GRASP framework and the experiment task. In addition, participants are provided with contacts for requesting further information or submitting complaints. The second section includes randomising the two scenarios and the two groups of predictive tools, to create the four scenarios described above.

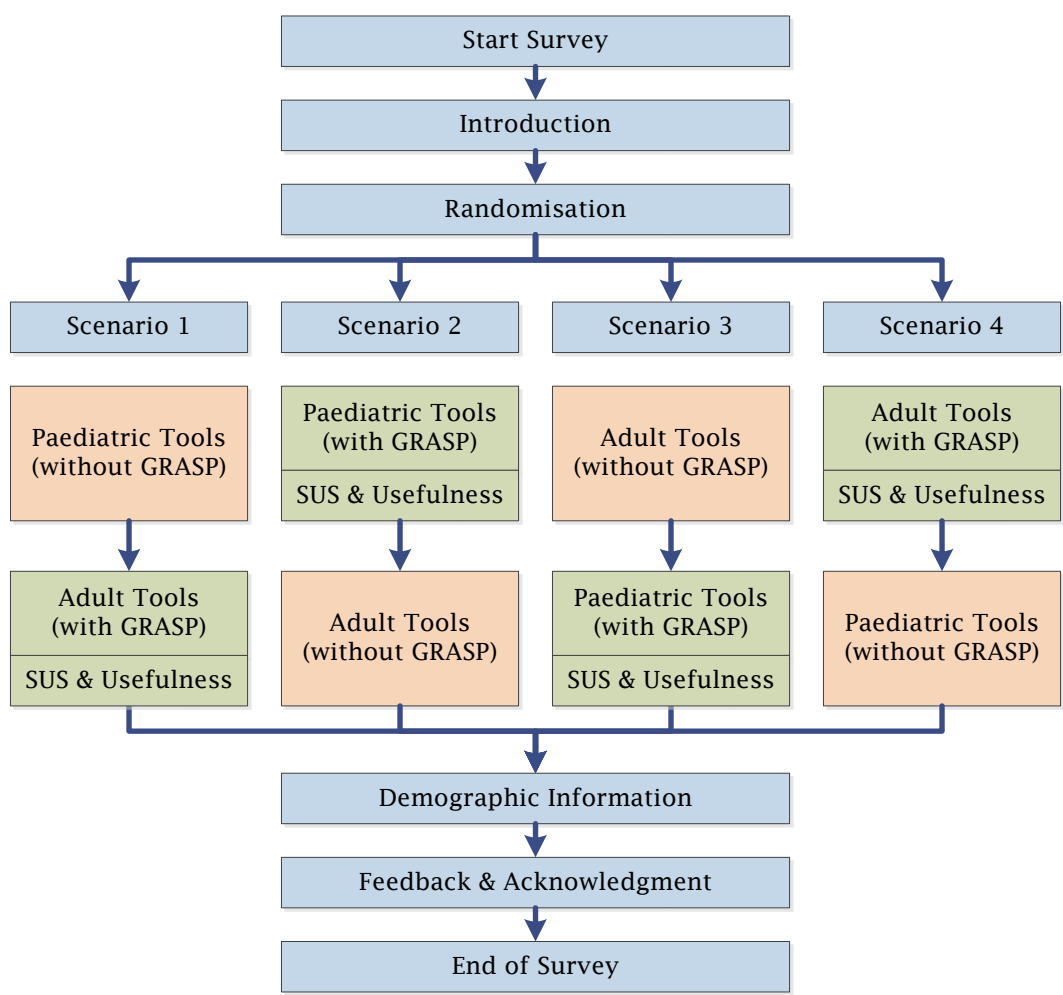

Figure 2: Survey Workflow and Randomisation of the Four Scenarios

In this second section, participants are asked to assume that they are the heads of busy emergency departments and they are responsible for selecting the best tool; the most validated in the literature or implemented in the clinical practice, out of five diagnostic head injury predictive tools. The PECARN is the correct answer among the five paediatric tools and both the CCHR and the NOC are correct answers among the five adult tools. On a five-point Likert scale, participants are asked to show how much they agree that; 1) they made their decisions based on guessing, 2) they made their decisions based on prior knowledge or experience, 3) they made their decisions based on the information provided in the survey, 4) they are confident in their decisions, and 5) they are satisfied with their decisions. The third section includes asking participants to provide their feedback on the usability of the evidence-based summary of the GRASP framework, through a standard set of System Usability Scale (SUS) questions [44, 45]. Participants were also asked to provide a free-text feedback on whether they think the framework is usefulness or not and why they think so. The fourth section includes participants' demographics, such as their clinical or healthcare role, specialty, gender, age group, years of experience, and how much they are familiar with head injury predictive tools.

### 2.3. The Study Sample Size

Based on similar studies, evaluating the impact of using information systems on professionals' objective, informed, and evidence-based decisions, the authors aimed to recruit a sample of 40 to 60 participants [46-48]. It was estimated that a sample size of 46 participants would be sufficient to test for differences in the measured outcomes when using a paired t-test with a significance level of 0.05 and power of 0.95. Calculations were conducted using G*Power version 3.1.9.2 [49].

### 2.4. Analysis and Outcomes

To test the five proposed hypotheses, the study was designed to compare the two scenarios; making decisions with and without using the GRASP framework, based on a group of seven measures. 1) Time needed for tools' selection decision making. 2) Accuracy of tools' selection decisions. 3) Making decisions subjectively based on guessing. 4) Making decisions subjectively based on prior knowledge and/or experience. 5) Making decisions objectively based on the information and evidence provided. 6) Levels of participants' confidence in their decisions. 7) Levels of participants' satisfaction with their decisions. The accuracy of making decisions, with and without GRASP, will also be compared along knowledge, experience, and familiarity with predictive tools. To avoid inflated Type I error and account for the five tested hypotheses, and the seven compared measures, the Bonferroni correction was used, through setting the alpha value of the paired samples t-test to 0.007 instead of 0.05. The sample size is re-estimated to 96 participants. In addition, the SUS was calculated for the average rate and the distribution of scores. The perceived usefulness and the free-text feedback were analysed. The demographic variables were analysed for possible correlations or differences.

## 3. Results

### *3.1. Descriptive Analysis*

Out of 5,857 relevant publications retrieved, a total of 3,282 professionals were identified and invited to take the survey. Over the survey duration of six weeks, from 11 March to 21 April 2019, a total of 194 valid responses were received, with a response rate of 5.9%. Valid responses were identified as those who completed the survey till the end and answered questions in all the survey sections, with no missing sections. Six participants missed answering one or more questions in one or more of the survey sections, five participants did not provide their demographics, and fifty-seven participants did not wish to be acknowledged in the study. The detailed distributions of participants based on gender, age group, years of experience, clinical and healthcare role, clinical specialty, familiarity with head injury predictive tools, and their countries are reported in the Appendix, in Tables 7 to 13 and illustrated in Figures 3 to 9.

### *3.2. The GRASP Impact on Participants' Decisions*

Using the GRASP framework; the evidence-based summary of predictive tools, the detailed report on each predictive tool, in addition to allowing participants to search the Internet, made them select the correct tools 88.1% of the time. Without GRASP; i.e. using the basic information about the predictive tools, the full text of the studies describing each tool, in addition to allowing participants to search the Internet, they selected the correct tools 53.7% of the time. This shows a statistically significant improvement of 64% (P<0.001). On a five-point Likert scale, where strongly agree is considered equal to five and strongly disagree is considered equal to one, the participants reported they made their tools' selection decisions based on guessing with an average of 1.98 when they used GRASP, compared to an average of 2.48 when they did not use GRASP. This shows a statistically significant reduction of 20% (P<0.001). Participants reported that they made their tools' selection decisions based on their prior knowledge and/or experience with an average of 3.27 when they used GRASP, compared to an average of 3.55 when they did not use GRASP. This shows a statistically significant reduction of 8% (P=0.0035).

Participants reported that they made their tools' selection decisions based on the provided information in the survey with an average of 4.10 when they used GRASP, compared to an average of 3.11 when they did not use GRASP. This shows a statistically significant increase of 32% (P<0.001). Participants reported that they were confident in their decisions with an average of 3.96 when they used GRASP, compared to an average of 3.55 when they did not use GRASP. This shows a statistically significant increase of 11% (P<0.001). Participants reported that they were satisfied with their decisions with an average of 3.99 when they used GRASP, compared to an average of 3.54 when they did not use GRASP. This shows a statistically significant increase of 13% (P<0.001). The duration of completing the task of selecting predictive tools showed high variability, with many statistical outliers. In addition to the average, the authors used the percentiles to avoid the effect of extreme outliers. The average duration of making the selection decisions showed a statistically insignificant reduction of 52% from 14.5 to 7.0 minutes (p= 0.385). There is also a reduction of 18.9% from 2.2 to 1.8 minutes on the 50th percentile, 37.3% from 5.3 to 3.3 minutes on the 75th percentile, 48.0% from 12.4 to 6.4 minutes on the 90th percentile, and 30.6% from 19.2 to 13.3 minutes on the 95th percentile. Table 1 shows the impact of using GRASP on the seven measures; decision accuracy, guessing, subjective decisions, objective decisions, confidence in decisions, satisfaction with decisions, and task completion duration 90th percentile in minutes.

Table 1: The Impact of Using GRASP on Participants' Decisions

| Criteria | Using GRASP | Score | Guessing | Subjective | Objective | Confidence | Satisfaction | Time in Min (90th Per) |
|---|---|---|---|---|---|---|---|---|
| **All Participants (n=194)** | No GRASP | 53.7% | 2.48 | 3.55 | 3.11 | 3.55 | 3.54 | 12.4 |
| | GRASP | 88.1% | 1.98 | 3.27 | 4.10 | 3.96 | 3.99 | 6.4 |
| | Change | 64% | -20% | -8% | 32% | 11% | 13% | -48% |
| | P-value | <0.0001 | <0.0001 | 0.0032 | <0.0001 | <0.0001 | <0.0001 | 0.3848 |

Using paired-samples t-test, Table 2 shows the estimation for paired difference of the seven measures and the effect size, calculating and interpreting the eta squared statistic, based on the guidelines propose by Cohen 2013 [50].

Table 2: Estimation for Paired Difference and Effect Size

| Measure | Mean | St Dev | SE Mean | 99.3% CI [1] | T-Value | P-Value | Effect Size [2] | |
|---|---|---|---|---|---|---|---|---|
| **Score** | 0.340 | 0.555 | 0.040 | (0.231, 0.449) | 8.53 | <0.0001 | 0.274 | Large |
| **Guessing** | -0.519 | 1.303 | 0.095 | (-0.777, -0.260) | -5.47 | <0.0001 | 0.134 | Moderate |
| **Subjective** | -0.319 | 1.464 | 0.107 | (-0.613, -0.028) | -2.99 | 0.0032 | 0.044 | Small |
| **Objective** | 1.005 | 1.496 | 0.109 | (0.709, 1.302) | 9.24 | <0.0001 | 0.307 | Large |
| **Confidence** | 0.392 | 1.261 | 0.092 | (0.141, 0.642) | 4.27 | <0.0001 | 0.086 | Moderate |
| **Satisfaction** | 0.439 | 1.235 | 0.090 | (0.194, 0.684) | 4.89 | <0.0001 | 0.110 | Moderate |
| **Duration** [3] | -447 | 7152 | 514 | (-1847, 952) | -0.87 | 0.3848 | Not Applicable | |

[1] Bonferroni correction conducted.
[2] Effect size calculated using eta squared statistic (0.01=small effect, 0.06=moderate effect, and 0.14=large effect [50])
[3] Task completion duration is reported in seconds.

Table 3 compares physicians to non-physicians, emergency medicine to other specialties, familiar with tools to non-familiar, males to females, younger to older and less experienced to more experienced participants. The GRASP detailed report is shown in Table 4 in the Appendix. The GRASP evidence-based summaries of the two groups of paediatric and adult predictive tools are shown in Tables 5 and 6 in the Appendix.

Table 3: Comparing the Impact of GRASP on Participants' Groups

| Criteria | Healthcare professional Group | Using GRASP | Score | Guessing | Subjective | Objective | Confidence | Satisfaction | Time in Min (90th Per) |
|---|---|---|---|---|---|---|---|---|---|
| Role | Physicians (n=130) | No GRASP | 61.4% | 2.4 | 3.7 | 3.0 | 3.6 | 3.6 | 10.9 |
| | | GRASP | 89.0% | 2.0 | 3.5 | 4.0 | 4.0 | 4.0 | 6.1 |
| | | Change | 45% | -18% | -5% | 33% | 10% | 12% | -44% |
| | | P-value | 0.000 | 0.000 | 0.080 | 0.000 | 0.001 | 0.000 | 0.620 |
| | Non-Physicians (n=59) | No GRASP | 37.3% | 2.7 | 3.3 | 3.5 | 3.5 | 3.5 | 15.3 |
| | | GRASP | 84.7% | 2.0 | 2.8 | 4.4 | 3.8 | 3.9 | 6.6 |
| | | Change | 127% | -25% | -16% | 28% | 10% | 14% | -57% |
| | | P-value | 0.000 | 0.000 | 0.007 | 0.000 | 0.047 | 0.008 | 0.263 |
| Specialty | Emergency (n=94) | No GRASP | 72.5% | 2.4 | 4.1 | 2.8 | 3.8 | 3.8 | 11.0 |
| | | GRASP | 93.4% | 1.9 | 3.7 | 3.8 | 4.1 | 4.1 | 6.5 |
| | | Change | 29% | -19% | -10% | 36% | 6% | 7% | -41% |
| | | P-value | 0.000 | 0.000 | 0.009 | 0.000 | 0.066 | 0.041 | 0.512 |
| | Non-Emergency (n=95) | No GRASP | 36.2% | 2.6 | 3.0 | 3.4 | 3.3 | 3.2 | 15.0 |
| | | GRASP | 83.0% | 2.0 | 2.9 | 4.4 | 3.8 | 3.8 | 6.5 |
| | | Change | 129% | -21% | -6% | 28% | 15% | 19% | -57% |
| | | P-value | 0.000 | 0.000 | 0.096 | 0.000 | 0.001 | 0.000 | 0.109 |
| Familiarity with tools | Familiar (n=108) | No GRASP | 67.0% | 2.3 | 4.1 | 2.8 | 3.8 | 3.8 | 8.1 |
| | | GRASP | 89.6% | 1.8 | 3.7 | 3.8 | 4.1 | 4.1 | 5.3 |
| | | Change | 34.0% | -22.0% | -10.0% | 39.0% | 8.0% | 8.0% | -34.0% |
| | | P-value | 0.000 | 0.000 | 0.007 | 0.000 | 0.016 | 0.013 | 0.512 |
| | Not Familiar (n=81) | No GRASP | 36.3% | 2.7 | 2.8 | 3.6 | 3.3 | 3.2 | 18.2 |
| | | GRASP | 85.0% | 2.2 | 2.7 | 4.5 | 3.7 | 3.8 | 7.9 |
| | | Change | 134% | -18% | -5% | 23% | 14% | 19% | -57% |
| | | P-value | 0.000 | 0.002 | 0.155 | 0.000 | 0.003 | 0.000 | 0.237 |
| Gender | Males (n=120) | No GRASP | 54.2% | 2.3 | 3.5 | 3.1 | 3.7 | 3.6 | 13.5 |
| | | GRASP | 82.2% | 2.0 | 3.3 | 4.1 | 3.9 | 4.0 | 7.4 |
| | | Change | 52% | -14% | -7% | 33% | 8% | 10% | -45% |
| | | P-value | 0.000 | 0.005 | 0.081 | 0.000 | 0.009 | 0.002 | 0.408 |
| | Females (n=67) | No GRASP | 54.5% | 2.9 | 3.5 | 3.3 | 3.3 | 3.4 | 12.2 |
| | | GRASP | 97.0% | 2.0 | 3.1 | 4.3 | 3.9 | 4.0 | 5.3 |
| | | Change | 78% | -30% | -12% | 29% | 17% | 18% | -56% |
| | | P-value | 0.000 | 0.000 | 0.004 | 0.000 | 0.004 | 0.001 | 0.536 |
| Age | Younger (<45 Years, n=112) | No GRASP | 58.7% | 2.6 | 3.6 | 3.1 | 3.5 | 3.5 | 9.1 |
| | | GRASP | 87.2% | 2.0 | 3.3 | 4.1 | 4.0 | 4.0 | 6.0 |
| | | Change | 48% | -25% | -7% | 34% | 13% | 14% | -34% |
| | | P-value | 0.000 | 0.000 | 0.057 | 0.000 | 0.001 | 0.001 | 0.446 |
| | Older (>45 Years, n=77) | No GRASP | 46.8% | 2.3 | 3.5 | 3.2 | 3.6 | 3.6 | 15.9 |
| | | GRASP | 88.3% | 2.0 | 3.2 | 4.1 | 3.9 | 4.0 | 7.7 |
| | | Change | 89% | -13% | -10% | 28% | 7% | 10% | -52% |
| | | P-value | 0.000 | 0.032 | 0.009 | 0.000 | 0.080 | 0.004 | 0.188 |
| Experience | Less Experience (<15 Years, n=94) | No GRASP | 58.7% | 2.6 | 3.6 | 3.0 | 3.5 | 3.4 | 8.1 |
| | | GRASP | 87.0% | 2.0 | 3.3 | 4.0 | 3.9 | 4.0 | 6.5 |
| | | Change | 48% | -24% | -7% | 36% | 12% | 16% | -20% |
| | | P-value | 0.000 | 0.000 | 0.090 | 0.000 | 0.009 | 0.001 | 0.460 |
| | More Experience (>15 Years, n=95) | No GRASP | 48.9% | 2.4 | 3.5 | 3.3 | 3.6 | 3.6 | 15.0 |
| | | GRASP | 88.3% | 2.0 | 3.2 | 4.2 | 3.9 | 4.0 | 6.8 |
| | | Change | 80% | -16% | -10% | 28% | 9% | 9% | -54% |
| | | P-value | 0.000 | 0.004 | 0.006 | 0.000 | 0.004 | 0.004 | 0.111 |

*3.3. The GRASP Usability and Usefulness*

The overall SUS rate of the GRASP framework and evidence-based summary, considering the responses of all 194 participants, was 72.5%, which represents a very good level of usability [51, 52]. Examining the influence of demographics on the SUS rates, only two factors showed significant influence; the gender of participants and their familiarity with predictive tools. Female participants reported a statistically significant higher SUS rate; 76.2%, compared to male participants; 70.8%, showing that female participants thought GRASP is easy to use more than male participants. Using statistical Spearman's correlation test; the degree of familiarity with head injury predictive tools showed a weak negative statistically significant correlation with the GRASP SUS score (p=0.031). This indicates that participants who were less familiar with predictive tools thought that the GRASP framework was easy to use more than participants who were more familiar with the tools.

Among the 194 valid responses of participants, almost two thirds; 122, provided a free-text feedback on the GRASP evidence-based summary usefulness and explained their feedback. Most participants, 88%, reported that they found the GRASP evidence-based summary useful. They explained their responses by various reasons; mainly that the evidence-based summary was simple, clear and logic. Some reported that the visual presentation was attractive, intuitive, and self-explanatory. Others reported that it provided a summary of much information in a concise and comprehensive way and some reported that the presented information was consistent, easily comparable, making informed decisions easier.

A smaller group of 12% of participants reported that they found the GRASP evidence-based summary not useful. They reported that it did not provide enough information to make informed decisions. Some reported that it was not clear enough, or simple enough, to understand and use to select predictive tools. One healthcare professional reported that "It is too complicated and needs to be simplified further", while another reported that "It is oversimplified and missing some important parameters". One healthcare professional reported "It might be more helpful when the decision is less clear" and added "I would like to see more info on the strengths/weaknesses of each tool".

## 4. Discussion and Conclusion

*4.1. Brief Summary*

It is a challenging task for most professionals to critically evaluate a growing number of predictive tools, proposed in the literature, in order to select effective tools for implementation at the clinical practice or for recommendation in clinical guidelines. Although most of these predictive tools have been assessed for predictive performance, only a few have been implemented and evaluated for comparative effectiveness or post-implementation impact. Professionals need an evidence-based approach to provide them with standardised objective information on predictive tools to support their search for and selection of effective tools for the clinical tasks. Based on the critical appraisal of the published evidence, the GRASP framework uses three dimensions to grade predictive tools: 1) Phase of Evaluation, 2) Level of Evidence and 3) Direction of Evidence. The final grade assigned to a tool is based on the highest phase of evaluation, supported by the highest level of positive evidence, or mixed evidence that supports a positive conclusion. In this manuscript, we present an evaluation of the impact of the GRASP framework on professionals' decisions in selecting predictive tools for clinical decision support.

*4.2. The GRASP Impact on Participants' Decisions*

The GRASP framework provides a systematic and transparent approach for professionals to make objective, well informed, and evidence-based decisions regarding selecting predictive tools. This is very similar to the findings of using the GRADE framework (Grading of Recommendations Assessment, Development and Evaluation) in evaluating the quality of evidence and strength of recommendations regarding treatment methods and decisions endorsed in clinical guidelines [53, 54]. The quality of decision making, while developing clinical guidelines, depends significantly on the quality of evidence-informed analysis and advice provided [55]. Similarly, supporting professionals with evidence improves their accuracy and help them make better clinical decisions as well as better organisational decisions [56, 57]. Likewise, using GRASP, and providing professionals with evidence-based information on predictive tools, significantly improved professionals' accuracy of decisions in selecting the best predictive tools.

Providing professionals with GRASP evidence-based information also enabled them to minimise subjective decision making, such as guessing, prior knowledge or previous experience. This has been proved in other studies discussing the role of utilising evidence-based resources in decreasing subjective bias in making clinical, population, and health policy decisions [58, 59]. Evidence-based information of GRASP proved also to decrease professionals' decisional conflict, through increasing their confidence in their decisions and their satisfaction with them. This has been proved in similar studies discussing the impact of evidence-based information on decreasing decisional conflicts faced by both professionals and patients when they make clinical decisions [60-62]. When time is a sensitive factor for critical clinical and population decisions, efficient decision making becomes important [63]. Here comes the role of evidence-based decision making, which proved to be not only more accurate, objective and of higher quality, but also much more efficient [64, 65]. Similarly, providing professionals with GRASP evidence-based information improved their efficiency in making predictive tools' selection decisions.

Using GRASP made nurses and other professionals make more accurate decisions than physicians, when they were not using GRASP. Using GRASP, clinicians of specialties other than emergency medicine made better decisions than emergency medicine clinicians without GRASP. Furthermore, using GRASP, professionals who were not familiar with head injury predictive tools made better decisions than professionals who were familiar with the tools without GRASP. Furthermore, using GRASP made decisions more efficient. Accordingly, using GRASP has moved professionals with less knowledge, less experience, and less familiarity with predictive tools to higher accuracy, higher efficiency, and better decision-making levels than professionals who had more knowledge, more experience, and were more familiar with tools, but were not using GRASP.

### *4.3. The GRASP Usability and Usefulness*

The usability of systems is an important foundation of their successful implementation and utilisation [66]. Usability can be evaluated through measuring the effectiveness of task management with accuracy and completeness, measuring efficiency of utilising resources in completing tasks and measuring users' satisfaction, comfort with, and positive attitudes towards, the use of the tools [67, 68]. One of the validated and simply applicable methods of measuring usability is the SUS [44, 45]. When users have more experience with a system they tend to provide higher, more favourable SUS scores for the system usability over users with either no or limited experience [69]. On the other hand, when users have less experience with a system, they tend to see new tools illustrating the system, or new approaches to understand it, more usable than users who have extensive experience with the system itself [70]. This explains why the degree of familiarity with the tools was negatively correlated with the GRASP SUS score, where participants less familiar with tools provided higher SUS scores for GRASP than participants who were more familiar. It is reported in the literature that gender has no influence on the perceived usability or usefulness of systems [71, 72]. This was not the case with GRASP, where female participants provided higher SUS scores than males. Furthermore, female participants also thought GRASP is useful more than males. Both findings could be explained by the greater improvement in female participants' confidence and satisfaction with their decisions, when they used GRASP, compared to male participants. Some participants' suggestions, reported in the free-text feedback, can be used in the future to add more information to the GRASP detailed report on each tool.

*4.4. Study Conclusion*

Through this study, the GRASP framework proved to be an effective evidence-based approach to support professionals' decisions when selecting predictive tools for implementation at the clinical practice or for recommendation in clinical guidelines. Using the GRASP framework and the evidence-based summary improved the accuracy of selecting the best predictive tools, with an increased objective, informed, and evidence-based decision making, and decreased subjective decision making; based on guessing, prior knowledge or experience. Using GRASP also decreased the decisional conflict facing professionals, through improving their confidence and satisfaction with their decisions. Using GRASP has also improved the efficiency of professionals in making their selection decisions, through decreasing the time needed to complete the decision-making task.

The GRASP framework represents a high-level approach to provide professionals with an evidence-based and comprehensive, yet simple and feasible, method to evaluate and select predictive tools. However, when professionals need further information, the framework detailed report provides them with the required details to support their decision making. The GRASP framework is not meant to be absolutely prescriptive. A lower grade tool could be preferred by a healthcare professional to improve clinical outcomes that are not supported by a higher grade one. It all depends on the objectives and priorities professionals are trying to achieve. More than one predictive tool could be endorsed, in clinical guidelines, each supported by its requirements and conditions of use and recommended for its most prominent outcomes of predictive performance or post-implementation impact on healthcare and clinical outcomes.

*4.5. Study Limitations and Future Work*

Even though we received a large and sufficient number of 194 valid responses, the very low response rate of 5.9% could have been improved if potential participants were motivated by some incentives. They could have also been motivated if more support was provided through their organisations, which needs more resources to synchronise such efforts. For the sake of keeping the survey feasible, for most busy professionals, the number of questions was kept limited and the required time to complete the survey was kept in the range of 20 minutes. However, some of the participants showed their willingness to provide more detailed feedback, which could have been done through interviews for example, but this was out of the scope of the study and was not initially possible with the huge number of invited participants. The reduction in the decision-making duration of selecting predictive tools, while using GRASP, was statistically insignificant, due to the high variability and extreme statistical outliers; with and without GRASP. This could be explained by the fact that the Qualtrics platform of the survey measures the task completion duration by subtracting the time of loading the page from the time of pushing the Next button after completing the task, and not the actual time the participants spent active on the page, which is currently under development [73].

To enable professionals and clinical guideline developers to access detailed information, reported evidence and assigned grades of predictive tools, it is essential to implement the GRASP framework into an online platform. However, maintaining such grading system up to date is a challenging task, as this requires continuous updating of the predictive tools grading and assessments, when new published evidence becomes available. It is essential to use automated or semi-automated methods for searching and processing new information to keep the GRASP framework information, grades, and assessments updated. Finally, we recommend that the GRASP framework be utilised by working groups of professional organisations to grade predictive tools, in order to provide consistent results and increase reliability and credibility for end users. These professional organisations should also support disseminating such evidence-based information on predictive tools, in a similar way of announcing and disseminating new updates of clinical practice guidelines.

## 5. Declarations

**Acknowledgment**

The authors would like to thank Peter Petocz, Professor of Statistics at Macquarie University, for his guidance and support in the statistical analysis. The authors would like also to thank all the professors, doctors, and researchers who participated in evaluating the impact of the GRASP framework, including: Aaron Brody, Adam Rose, Adel Hamed Elbaih, Aleix Martínez-Pérez, Alireza Razzaghi, Amanda Montalbano, Andreas Huefner, Andreas Sundermeyer, Andres M Rubiano, Andrew MacCormick, Andrew Miller, Antoinette Conca, Asit Misra, Ausra Snipaitiene, Azadeh Tafakori, Başak Bayram, Bedia Gulen, Bhagavatula Indira Devi, Brent Becker, Brian Rowe, Chad Cannon, Chen-June Seak, Chimwemwe Mula, Christian Waydhas, Christophe Marti, Christopher R. Carpenter, Claire Vajdic, Claudine Blum, Clermont E. Dionne, Corrado Tagliati, David R. Vinson, Dawn Moeller, Deepak Batura, Deepika Mohan, DJ Choi, Duane Steward, Dustin Ballard, Edward Baker, Elise Gane, Elizabeth Manias, Elizabeth Powell, Ellen Weber, Emanuele Gilardi, Eveline Hitti, Ewout W. Steyerberg, Eva Bitzer, Fawaz Abdullah Alharbi, Fayza Haider, Fernanda Bellolio, Fleur Lorton, Francesco Fleres, Gabriel Rodrigues, Gerry Lee, Giacomo Veronese, Gianfranco Sinagra, Giovanni de Simone, Guillaume Foldes-Busque, Gulten Sucu, Guo XH, Gwendolyn Vuurberg, Herbert J. Fernandes, Hon Lon Tam, Hsu Teh-Fu, Ibrahim Jatau Abubakar, Ilias Galanopoulos, Iqra Manzoor, Jennifer Hoffmann, JM Ferreras, Joel Noutakdie Tochie, John Kellett, Jonathan Elmer, Jorge Acosta-Reyes, Juan Del Castillo-Calcáneo, Julia Lappin, Kai Eggers, Karin Pukk Härenstam, Katarina Lockman Frostred, Ken Hillman, Kevin Heard, Kimberly D. Johnson, Kristen Miller, Kristin Baltrusaitis, L Beenen, Larry Figgs, Lauren Southerland, Lina Jankauskaitė, Luca Molinari, Majed Althagafi, Makini Chisolm-Straker, Marc Probst, Marcello Covino, Marco Daverio, Marie Méan, Mariza Elsi, Mark Ebell, Martin Nordberg, Matt Reed, Matthew J Douma, Mauro Podda, Melanie Dechamps, Michael Joseph Barrett, Mieke Deschodt, Mojtaba Mojtahedzadeh, Molly Moore Jeffery, Muhammad Waseem, Nathan Hoot, Niccolò Parri, Nicola Ramacciati, Nurhafiza Yezid, Olga H Torres, Özlem Köksal, Paola Fugazzola, Paolo Navalesi, Paul L. Aronson, Paul Monagle, Pedro J Marcos, Peter Dayan, Peter Nugus, Pinchas Halpern, Prabath Lodewijks, Prosen Gregor, Rachel Seymour, Rachid Mahmoudi, Rafael Oliveira Ximenes, Rafal Karwowski, Rahul Kashyap, Ricardo Fernandes, Rodolfo J. Oviedo, Ross I. Donaldson, Sabrina De Winter, Sandeep Sahu, Sangil Lee, Sebastián Camino, Sheri Carson, Sivera Berben, Stephane Tshitenge, Sukhyang Lee, Suzanne Tamang, Thomas Hartka, Thys Frédéric, Tim Söderlund, Tiziana Ciarambino, Toby Keene, Tomas Vedin, Vincenzo G. Menditto, Wei-Chieh Lee, William Mower.

**Funding**

This work was supported by the Commonwealth Government Funded Research Training Program, Australia.

**Authors' contributions**

MK mainly contributed to the conception, detailed design, and conduction of the study. BG and FM supervised the study from the scientific perspective. BG was responsible for the overall supervision of the work done, while FM was responsible for providing advice on the enhancement of the methodology used. All the authors have been involved in drafting the manuscript and revising it. Finally, all the authors gave approval of the manuscript to be published and agreed to be accountable for all aspects of the work.

**Ethics approval and consent to participate**

This study has been approved by the Human Research Ethics Committee, Faculty of Medicine and Health Sciences, Macquarie University, Sydney, Australia, on the 4$^{th}$ of October 2018. Reference No: 5201834324569. Project ID: 3432.

**Consent to publication**

Not applicable.

**Competing interests**

The authors declare that they have no competing interests.

**Author details**

[1] Australian Institute of Health Innovation, Faculty of Medicine and Health Sciences, Macquarie University, 75 Talavera Road, North Ryde, Sydney, NSW 2113, Australia.

[2] Centre for Big Data Research in Health, University of New South Wales, Kensington, Sydney, NSW 2052, Australia.

# 6. References

1. Chaudhry, B., et al., *Systematic review: impact of health information technology on quality, efficiency, and costs of medical care.* Ann Intern Med, 2006. **144**(10): p. 742-52.
2. Garg, A.X., et al., *Effects of computerized clinical decision support systems on practitioner performance and patient outcomes: a systematic review.* Jama, 2005. **293**(10): p. 1223-1238.
3. Kawamoto, K., et al., *Improving clinical practice using clinical decision support systems: a systematic review of trials to identify features critical to success.* Bmj, 2005. **330**(7494): p. 765.
4. Oman, K.S., *Evidence-based practice: An implementation guide for healthcare organizations.* 2010: Jones & Bartlett Publishers.
5. Osheroff, J.A. *Improving outcomes with clinical decision support: an implementer's guide.* 2012. Himss.
6. Osheroff, J.A., et al., *A roadmap for national action on clinical decision support.* Journal of the American medical informatics association, 2007. **14**(2): p. 141-145.
7. Musen, M.A., B. Middleton, and R.A. Greenes, *Clinical decision-support systems*, in *Biomedical informatics.* 2014, Springer. p. 643-674.
8. Shortliffe, E.H. and J.J. Cimino, *Biomedical informatics: computer applications in health care and biomedicine.* 2013: Springer Science & Business Media.
9. Adams, S.T. and S.H. Leveson, *Clinical prediction rules.* Bmj, 2012. **344**: p. d8312.
10. Wasson, J.H., et al., *Clinical prediction rules: applications and methodological standards.* New England Journal of Medicine, 1985. **313**(13): p. 793-799.
11. Beattie, P. and R. Nelson, *Clinical prediction rules: what are they and what do they tell us?* Australian Journal of Physiotherapy, 2006. **52**(3): p. 157-163.
12. Steyerberg, E.W., *Clinical prediction models: a practical approach to development, validation, and updating.* 2008: Springer Science & Business Media.
13. Ebell, M.H., *Evidence-based diagnosis: a handbook of clinical prediction rules.* Vol. 1. 2001: Springer Science & Business Media.
14. Kappen, T., et al., *General Discussion I: Evaluating the Impact of the Use of Prediction Models in Clinical Practice: Challenges and Recommendations.* Prediction Models and Decision Support, 2015: p. 89.
15. Taljaard, M., et al., *Cardiovascular Disease Population Risk Tool (CVDPoRT): predictive algorithm for assessing CVD risk in the community setting. A study protocol.* BMJ open, 2014. **4**(10): p. e006701.
16. Dictionary, M.-W. *Definition of clinician.* 2019 [cited 2019; Available from: https://www.merriam-webster.com/dictionary/clinician.
17. Ansari, S. and A. Rashidian, *Guidelines for guidelines: are they up to the task? A comparative assessment of clinical practice guideline development handbooks.* PloS one, 2012. **7**(11): p. e49864.

18. Kish, M.A., *Guide to development of practice guidelines.* Clinical Infectious Diseases, 2001. **32**(6): p. 851-854.
19. Shekelle, P.G., et al., *Developing clinical guidelines.* Western Journal of Medicine, 1999. **170**(6): p. 348.
20. Tranfield, D., D. Denyer, and P. Smart, *Towards a methodology for developing evidence-informed management knowledge by means of systematic review.* British journal of management, 2003. **14**(3): p. 207-222.
21. Raine, R., C. Sanderson, and N. Black, *Developing clinical guidelines: a challenge to current methods.* Bmj, 2005. **331**(7517): p. 631.
22. O'Connor, A., *User manual-decisional conflict scale.* 2010. URL: http://decisionaid. ohri. ca/docs/develop/User_Manuals/UM_Decisional_Conflict. pdf, 2017.
23. Khalifa, M., F. Magrabi, and B. Gallego, *Developing an Evidence-Based Framework for Grading and Assessment of Predictive Tools for Clinical Decision Support.* arXiv preprint arXiv:1907.03706, 2019.
24. Azim, A. and B. Joseph, *Traumatic brain injury*, in *Surgical Critical Care Therapy.* 2018, Springer. p. 1-10.
25. Greve, M.W. and B.J. Zink, *Pathophysiology of traumatic brain injury.* Mount Sinai Journal of Medicine: A Journal of Translational and Personalized Medicine: A Journal of Translational and Personalized Medicine, 2009. **76**(2): p. 97-104.
26. Maguire, J.L., et al., *Should a head-injured child receive a head CT scan? A systematic review of clinical prediction rules.* Pediatrics, 2009. **124**(1): p. e145-e154.
27. Maguire, J.L., et al., *Clinical prediction rules for children: a systematic review.* Pediatrics, 2011: p. peds. 2011-0043.
28. Atabaki, S.M., et al., *A clinical decision rule for cranial computed tomography in minor pediatric head trauma.* Archives of pediatrics & adolescent medicine, 2008. **162**(5): p. 439-445.
29. Osmond, M.H., et al., *CATCH: a clinical decision rule for the use of computed tomography in children with minor head injury.* Canadian Medical Association Journal, 2010. **182**(4): p. 341-348.
30. Dunning, J., et al., *Derivation of the children's head injury algorithm for the prediction of important clinical events decision rule for head injury in children.* Archives of disease in childhood, 2006. **91**(11): p. 885-891.
31. Kuppermann, N., et al., *Identification of children at very low risk of clinically-important brain injuries after head trauma: a prospective cohort study.* The Lancet, 2009. **374**(9696): p. 1160-1170.
32. Palchak, M.J., et al., *A decision rule for identifying children at low risk for brain injuries after blunt head trauma.* Annals of emergency medicine, 2003. **42**(4): p. 492-506.

33. Bressan, S., et al., *Implementation of adapted PECARN decision rule for children with minor head injury in the pediatric emergency department.* Academic Emergency Medicine, 2012. **19**(7): p. 801-807.
34. Atabaki, S.M., et al., *Quality Improvement in Pediatric Head Trauma with PECARN Rules Implementation as Computerized Decision Support.* Pediatric Quality & Safety, 2017. **2**(3): p. e019.
35. Stiell, I.G., et al., *The Canadian CT Head Rule for patients with minor head injury.* The Lancet, 2001. **357**(9266): p. 1391-1396.
36. Bezuidenhout, A.F., et al., *The Kimberley Hospital Rule (KHR) for urgent computed tomography of the brain in a resource-limited environment.* South African Medical Journal, 2013. **103**(9): p. 646-651.
37. Miller, E.C., J.F. Holmes, and R.W. Derlet, *Utilizing clinical factors to reduce head CT scan ordering for minor head trauma patients.* Journal of Emergency Medicine, 1997. **15**(4): p. 453-457.
38. Ibañez, J., et al., *Reliability of clinical guidelines in the detection of patients at risk following mild head injury: results of a prospective study.* Journal of neurosurgery, 2004. **100**(5): p. 825-834.
39. Haydel, M.J., et al., *Indications for computed tomography in patients with minor head injury.* New England Journal of Medicine, 2000. **343**(2): p. 100-105.
40. Bouida, W., et al., *Prediction value of the Canadian CT head rule and the New Orleans criteria for positive head CT scan and acute neurosurgical procedures in minor head trauma: a multicenter external validation study.* Annals of emergency medicine, 2013. **61**(5): p. 521-527.
41. Papa, L., et al., *Performance of the Canadian CT Head Rule and the New Orleans Criteria for predicting any traumatic intracranial injury on computed tomography in a United States Level I trauma center.* Academic Emergency Medicine, 2012. **19**(1): p. 2-10.
42. Smits, M., et al., *External validation of the Canadian CT Head Rule and the New Orleans Criteria for CT scanning in patients with minor head injury.* Jama, 2005. **294**(12): p. 1519-1525.
43. Qualtrics Experience Management Solutions, Q. *Qualtrics Experience Management Solutions, Qualtrics.* 2018 [cited 2018 1 January]; Available from: https://www.qualtrics.com/.
44. Brooke, J., *SUS-A quick and dirty usability scale.* Usability evaluation in industry, 1996. **189**(194): p. 4-7.
45. Brooke, J., *SUS: a retrospective.* Journal of usability studies, 2013. **8**(2): p. 29-40.
46. Del Fiol, G., et al., *Effectiveness of topic-specific infobuttons: a randomized controlled trial.* Journal of the American Medical Informatics Association, 2008. **15**(6): p. 752-759.

47. Schardt, C., et al., *Utilization of the PICO framework to improve searching PubMed for clinical questions.* BMC medical informatics and decision making, 2007. **7**(1): p. 16.
48. Westbrook, J.I., E.W. Coiera, and A.S. Gosling, *Do online information retrieval systems help experienced clinicians answer clinical questions?* Journal of the American Medical Informatics Association, 2005. **12**(3): p. 315-321.
49. Faul, F., et al., *G* Power 3: A flexible statistical power analysis program for the social, behavioral, and biomedical sciences.* Behavior research methods, 2007. **39**(2): p. 175-191.
50. Cohen, J., *Statistical power analysis for the behavioral sciences.* 2013: Routledge.
51. Bangor, A., P. Kortum, and J. Miller, *Determining what individual SUS scores mean: Adding an adjective rating scale.* Journal of usability studies, 2009. **4**(3): p. 114-123.
52. Kortum, P.T. and A. Bangor, *Usability ratings for everyday products measured with the System Usability Scale.* International Journal of Human-Computer Interaction, 2013. **29**(2): p. 67-76.
53. Alonso-Coello, P., et al., *GRADE Evidence to Decision (EtD) frameworks: a systematic and transparent approach to making well informed healthcare choices. 1: Introduction.* bmj, 2016. **353**: p. i2016.
54. Guyatt, G.H., et al., *GRADE: an emerging consensus on rating quality of evidence and strength of recommendations.* Bmj, 2008. **336**(7650): p. 924-926.
55. Head, B.W., *Toward more "evidence-informed" policy making?* Public Administration Review, 2016. **76**(3): p. 472-484.
56. Barends, E. and D.M. Rousseau, *Evidence-based management: How to use evidence to make better organizational decisions.* 2018: Kogan Page Publishers.
57. Castaneda, C., et al., *Clinical decision support systems for improving diagnostic accuracy and achieving precision medicine.* Journal of clinical bioinformatics, 2015. **5**(1): p. 4.
58. Hunink, M.M., et al., *Decision making in health and medicine: integrating evidence and values.* 2014: Cambridge University Press.
59. McCaughey, D. and N.S. Bruning, *Rationality versus reality: the challenges of evidence-based decision making for health policy makers.* Implementation Science, 2010. **5**(1): p. 39.
60. Bate, L., et al., *How clinical decisions are made.* British journal of clinical pharmacology, 2012. **74**(4): p. 614-620.
61. McAlpine, K., et al., *What is the effectiveness of patient decision aids for cancer-related decisions? A systematic review subanalysis.* JCO clinical cancer informatics, 2018. **2**: p. 1-13.
62. Vlemmix, F., et al., *Decision aids to improve informed decision-making in pregnancy care: a systematic review.* BJOG: An International Journal of Obstetrics & Gynaecology, 2013. **120**(3): p. 257-266.

63. Brownson, R.C., et al., *Evidence-based public health.* 2017: Oxford University Press.
64. Mickan, S., et al., *Evidence of effectiveness of health care professionals using handheld computers: a scoping review of systematic reviews.* Journal of medical Internet research, 2013. **15**(10): p. e212.
65. Ventola, C.L., *Mobile devices and apps for health care professionals: uses and benefits.* Pharmacy and Therapeutics, 2014. **39**(5): p. 356.
66. Delone, W.H. and E.R. McLean, *The DeLone and McLean model of information systems success: a ten-year update.* Journal of management information systems, 2003. **19**(4): p. 9-30.
67. Frøkjær, E., M. Hertzum, and K. Hornbæk. *Measuring usability: are effectiveness, efficiency, and satisfaction really correlated?* in *Proceedings of the SIGCHI conference on Human Factors in Computing Systems.* 2000. ACM.
68. Khajouei, R., et al., *Clinicians satisfaction with CPOE ease of use and effect on clinicians' workflow, efficiency and medication safety.* international journal of medical informatics, 2011. **80**(5): p. 297-309.
69. McLellan, S., A. Muddimer, and S.C. Peres, *The effect of experience on System Usability Scale ratings.* Journal of usability studies, 2012. **7**(2): p. 56-67.
70. Albert, W. and T. Tullis, *Measuring the user experience: collecting, analyzing, and presenting usability metrics.* 2013: Newnes.
71. Harrati, N., et al., *Exploring user satisfaction for e-learning systems via usage-based metrics and system usability scale analysis.* Computers in Human Behavior, 2016. **61**: p. 463-471.
72. Orfanou, K., N. Tselios, and C. Katsanos, *Perceived usability evaluation of learning management systems: Empirical evaluation of the System Usability Scale.* The International Review of Research in Open and Distributed Learning, 2015. **16**(2).
73. Barnhoorn, J.S., et al., *QRTEngine: An easy solution for running online reaction time experiments using Qualtrics.* Behavior research methods, 2015. **47**(4): p. 918-929.

# 7. The Appendix

## 7.1. The GRASP Framework Detailed Report

Table 4: The GRASP Framework Detailed Report

| Name | Name of predictive tool (report tool's creators and year in the absence of a given name) |
|---|---|
| Author | Name of developer (first author or researcher) |
| Country | Country of development |
| Year | Year of development |
| Category | Diagnostic/Therapeutic/Prognostic/Preventive |
| Intended use | Specific aim/intended use of the predictive tool |
| Intended user | Type of practitioner intended to use the tool |
| Clinical area | Clinical specialty |
| Target Population | Target patient population and health care settings in which the tool is applied |
| Target Outcome | Event to be predicted (including prediction lead time if needed) |
| Action | Recommended action based on tool's output |
| Input source | • Clinical (including Diagnostic, Genetic, Vital signs, Pathology)<br>• Non-Clinical (including Healthcare Utilisation) |
| Input type | • Objective (Measured input; from electronic systems or clinical examination)<br>• Subjective (Patient reported; history, checklist …etc.) |
| Local context | Is the tool developed using location-specific data? (e.g. life expectancy tables) |
| Methodology | Type of algorithm used for developing the tool (e.g. parametric/non-parametric) |
| Internal Validation | Method of internal validation |
| Dedicated Support | Name of the supporting/funding research networks, programs, or professional groups |
| Endorsement | Organisations endorsing the tool and/or clinical guidelines recommending its utilisation |
| Automation Flag | Automation status (manual/automated) |
| Tool Citations | Total citations of the tool |
| Studies | Number of studies reporting the tool |
| Authors No | Number of authors |
| Sample Size | Size of patient/record sample used in the development of the tool |
| Journal Name | Name of the journal that published the tool's primary development study |
| Journal Rank | Impact factor of the journal |
| Citation Index | Calculated as: Average Annual Citations = number of citations/age of primary publication |
| Publication Index | Calculated as: Average Annual Studies = number of studies/age of primary publication |
| Literature Index | Calculated as: Citations and Publications = number of citations X number of studies |

| Phase of Evaluation | Level of Evidence | Grade | Evaluation Studies |
|---|---|---|---|
| Phase C:<br><br>Before implementation<br><br>Is it possible? | Insufficient internal validation | C0 | Not tested for internal validity, insufficiently internally validated, or internal validation was insufficiently reported. |
| | Internal validation | C3 | Tested for internally validity (reported calibration & discrimination; sensitivity, specificity, positive and negative predictive values & other predictive performance measures). |
| | External validation | C2 | Tested for external validity, using one external dataset. |
| | External validation multiple times | C1 | Tested multiple times for external validity, using more than one external dataset. |

| Phase B:<br>Planning for implementation<br>Is it practicable? | Usability | B3 | Reported usability testing (tool effectiveness, efficiency, satisfaction, learnability, memorability, and minimizing errors). | | | | | | |
|---|---|---|---|---|---|---|---|---|---|
| | Potential effect | B2 | Reported estimated potential effect on clinical effectiveness, patient safety or healthcare efficiency. | | | | | | |
| | Potential effect & Usability | B1 | Both potential effect and usability are reported. | | | | | | |
| Phase A:<br>After implementation:<br>Is it desirable? | Evaluation of post-implementation impact on Clinical Effectiveness, Patient Safety or Healthcare Efficiency | A3 | Based on subjective studies; e.g. the opinion of a respected authority, clinical experience, a descriptive study, or a report of an expert committee or panel. | | | | | | |
| | | A2 | Based on observational studies; e.g. a well-designed cohort or case-control study. | | | | | | |
| | | A1 | Based on experimental studies; properly designed, widely applied randomised/nonrandomised controlled trial. | | | | | | |
| Assigned Grade | Grade ABC/123 | A1 | A2 | A3 | B1 | B2 | B3 | C1 | C2 | C3 |
| Direction of Evidence | ● Positive Evidence | ◐ Mixed Evidence Supporting Positive Conclusion | | | | | | | | |
| | ○ Negative Evidence | ◑ Mixed Evidence Supporting Negative Conclusion | | | | | | | | |
| Justification | Explains how the final grade is assigned based on evidence; which conclusions were taken into consideration, as positive evidence, and which were considered negative. | | | | | | | | | |
| References | Details of studies that support the justification: phase of evaluation, level of evidence, direction of evidence, study type, study settings, methodology, results, findings and conclusions (highlighted according to the findings codes). | | | | | | | | | |
| Findings Codes | Positive Findings / Negative Findings / Important Findings | | | | | | | | | |

## 7.2. The Paediatric and Adult Head Injury Predictive Tools

Table 5: The GRASP Evidence-Based Summary of Paediatric Head Injury Predictive Tools

| Tool | Tool Information | | | | Tool Grade | Impact After Implementation | | | Planning for Implementation | | | Performance Before Implementation | | |
|---|---|---|---|---|---|---|---|---|---|---|---|---|---|---|
| | Country | Year | Citations | Studies | | Experimental Studies | Observational Studies | Subjective Studies | Potential Effect & Usability | Potential Effect | Usability | External Validation Multiple Times | External Validation Only Once | Internal Validation |
| | | | | | | A1 | A2 | A3 | B1 | B2 | B3 | C1 | C2 | C3 |
| PECARN | USA | 2009 | 886 | 24 | A2 | | ◐ (Mixed Positive) | | | ● | | ● | | ● |
| CHALICE | UK | 2006 | 308 | 15 | B2 | | | | | ◐ (Mixed Negative) | | ● | | ● |
| CATCH | USA | 2006 | 321 | 12 | C1 | | | | | | | ● | | ● |
| Palchak | USA | 2003 | 247 | 3 | C2 | | | | | | | | ● | ● |
| Atabaki | USA | 2008 | 111 | 1 | C3 | | | | | | | | | ● |
| Evidence Direction | ● Positive Evidence    ○ Negative Evidence | | | | | ◐ Mixed Evidence Supporting Positive Conclusion    ◐ Mixed Evidence Supporting Negative Conclusion | | | | | | | | |

Table 6: The GRASP Evidence-Based Summary of Adult Head Injury Predictive Tools

| Tool | Tool Information | | | | Tool Grade | Impact After Implementation | | | Planning for Implementation | | | Performance Before Implementation | | |
|---|---|---|---|---|---|---|---|---|---|---|---|---|---|---|
| | Country | Year | Citations | Studies | | Experimental Studies | Observational Studies | Subjective Studies | Potential Effect & Usability | Potential Effect | Usability | External Validation Multiple Times | External Validation Only Once | Internal Validation |
| | | | | | | A1 | A2 | A3 | B1 | B2 | B3 | C1 | C2 | C3 |
| CCHR | Canada | 2001 | 1098 | 23 | C1 | ○ | ○ | | | ○ | | ● | | ● |
| NOC | USA | 2000 | 907 | 11 | C1 | | | | | ○ | | ● | | ● |
| Miller | USA | 1997 | 210 | 2 | C3 | | | | | | | | ○ | ● |
| KHR | S Africa | 2013 | 7 | 1 | C3 | | | | | | | | | ● |
| Ibanez | Spain | 2004 | 165 | 1 | C0 | | | | | | | | | ○ |
| Evidence Direction | ● Positive Evidence    ○ Negative Evidence | | | | | ◐ Mixed Evidence Supporting Positive Conclusion    ◐ Mixed Evidence Supporting Negative Conclusion | | | | | | | | |

## 7.3. Statistical Tables and Figures

Table 7: Gender Distribution of Participants

| Gender | Count | Percentage | Cum |
|---|---|---|---|
| Males | 120 | 61.9% | 61.9% |
| Females | 67 | 34.5% | 96.4% |
| Not Reported | 7 | 3.6% | 100.0% |
| Total | **194** | **100%** | |

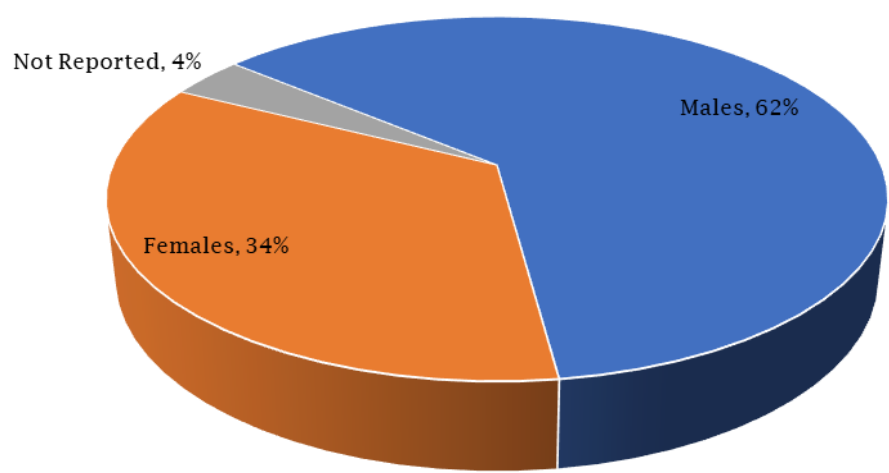

Figure 3: Gender Distribution of Participants

Table 8: Age Group Distribution of Participants

| Age Group | Count | Percentage | Cum |
|---|---|---|---|
| 25 – 34 years | 23 | 11.9% | 11.9% |
| 35 - 44 years | 89 | 45.9% | 57.7% |
| 45 - 54 years | 49 | 25.3% | 83.0% |
| 55 - 64 years | 20 | 10.3% | 93.3% |
| 65 - 74 years | 7 | 3.6% | 96.9% |
| 75 years or older | 1 | 0.5% | 97.4% |
| Not Reported | 5 | 2.6% | 100.0% |
| **Total** | **194** | **100%** | |
| **Average Age** | **44.8 years** | | |

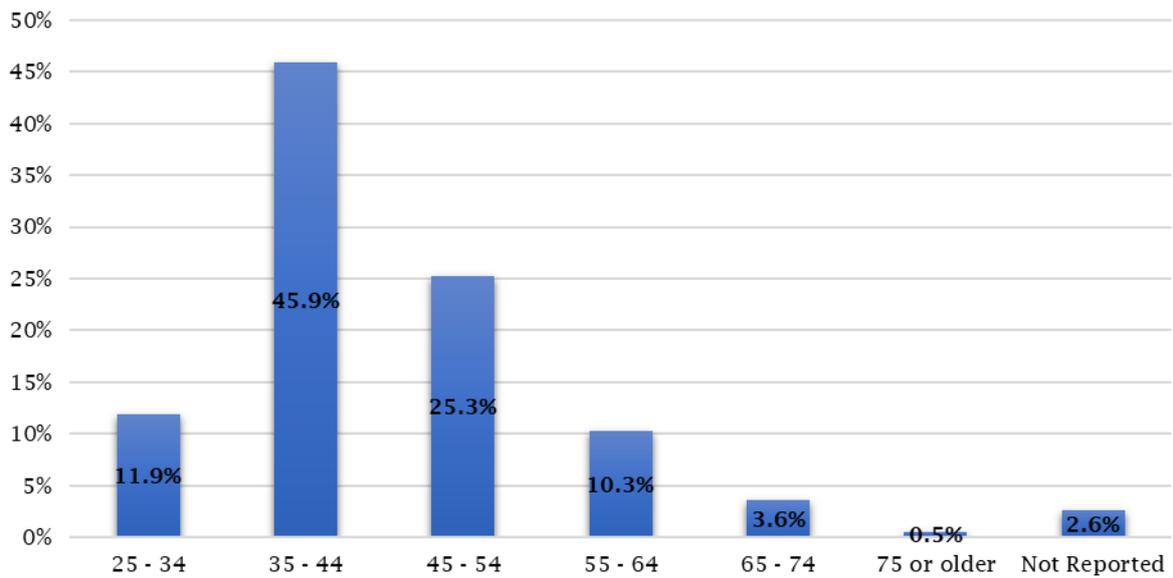

Figure 4: Age Group Distribution of Participants

Table 9: Years of Experience Distribution of Participants

| Years of Experience | Count | Percentage | Cum |
|---|---|---|---|
| Less than 5 years | 12 | 6.2% | 6.2% |
| 05-09 years | 29 | 14.9% | 21.1% |
| 10-14 years | 53 | 27.3% | 48.5% |
| 15-19 years | 32 | 16.5% | 64.9% |
| 20-24 years | 22 | 11.3% | 76.3% |
| 25-29 years | 19 | 9.8% | 86.1% |
| 30-34 years | 12 | 6.2% | 92.3% |
| 35 years or more | 10 | 5.2% | 97.4% |
| Not Reported | 5 | 2.6% | 100.0% |
| **Total** | **194** | **100%** | |
| **Average Age** | **16.7 years** | | |

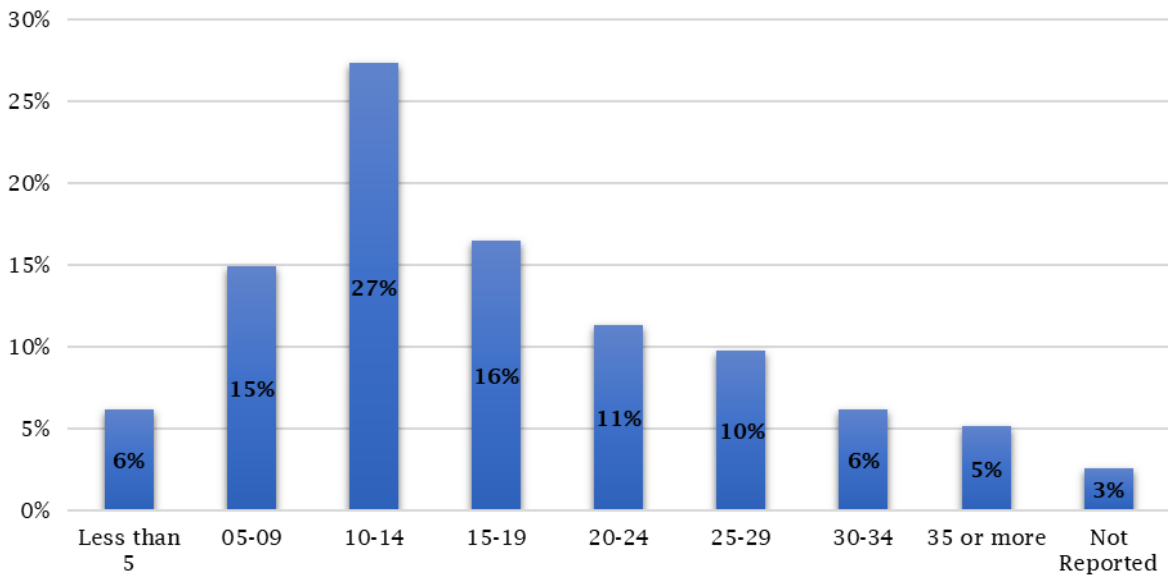

Figure 5: Years of Experience Distribution of Participants

Table 10: Role Distribution of Participants

| Clinical Role | Count | Percentage | Cum |
|---|---|---|---|
| Physician - Consultant | 114 | 58.8% | 58.8% |
| Physician - Registrar | 10 | 5.2% | 63.9% |
| Physician - Resident | 6 | 3.1% | 67.0% |
| Nurse - Senior | 13 | 6.7% | 73.7% |
| Nurse - Junior | 0 | 0.0% | 73.7% |
| Other Healthcare professionals | 46 | 23.7% | 97.4% |
| Not Reported | 5 | 2.6% | 100.0% |
| **Total** | **194** | **100%** | |

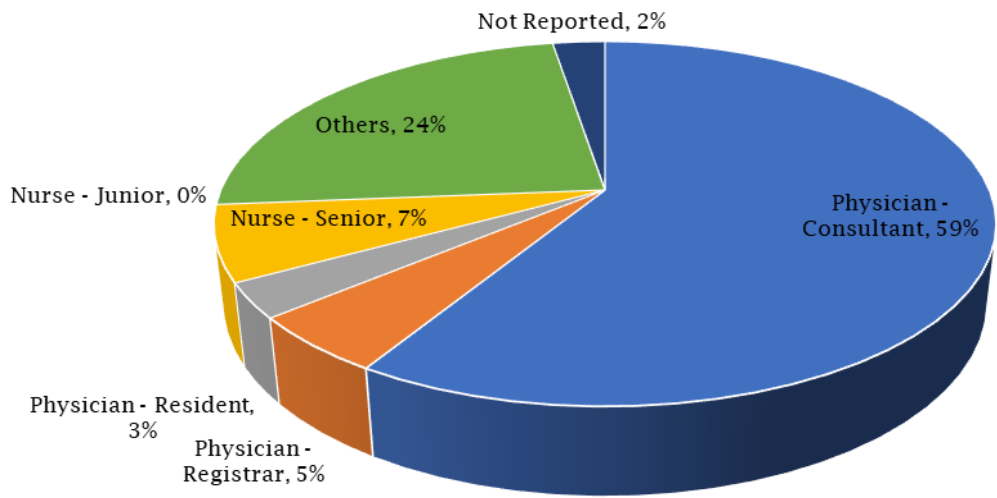

Figure 6: Role Distribution of Participants

Table 11: Specialty Distribution of Participants

| Clinical Specialty | Count | Percentage | Cum |
|---|---|---|---|
| Emergency Medicine | 94 | 48.5% | 48.5% |
| Other Specialties | 95 | 49.0% | 97.4% |
| Not Reported | 5 | 2.6% | 100.0% |
| **Total** | **194** | **100%** | |

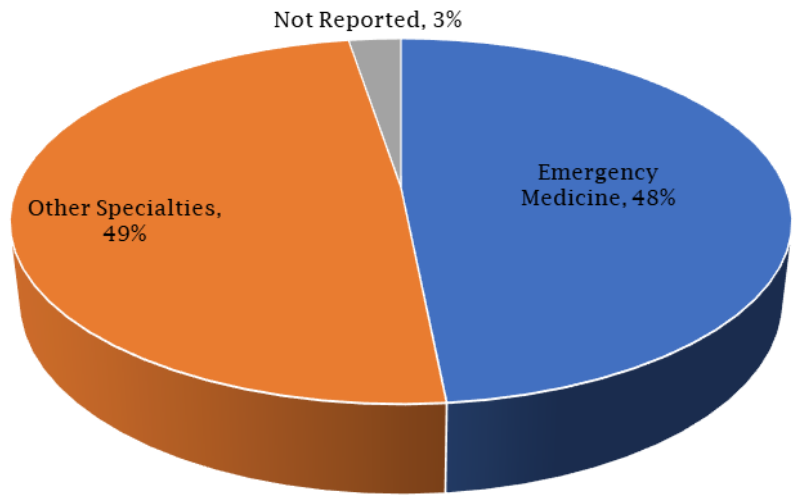

Figure 7: Specialty Distribution of Participants

Table 12: Participants are Familiar with Predictive Tools

| Familiar with Tools | Count | Percentage | Cum |
|---|---|---|---|
| Strongly Agree (code = 5) | 47 | 24.2% | 24.2% |
| Somewhat Agree (code = 4) | 61 | 31.4% | 55.7% |
| Neither Agree nor Disagree (code = 3) | 18 | 9.3% | 64.9% |
| Somewhat Disagree (code = 2) | 33 | 17.0% | 82.0% |
| Strongly Disagree (code = 1) | 30 | 15.5% | 97.4% |
| Not Reported | 5 | 2.6% | 100.0% |
| Total | 194 | 100% ||
| Average | Neither Agree nor Disagree (score = 3.33) |||

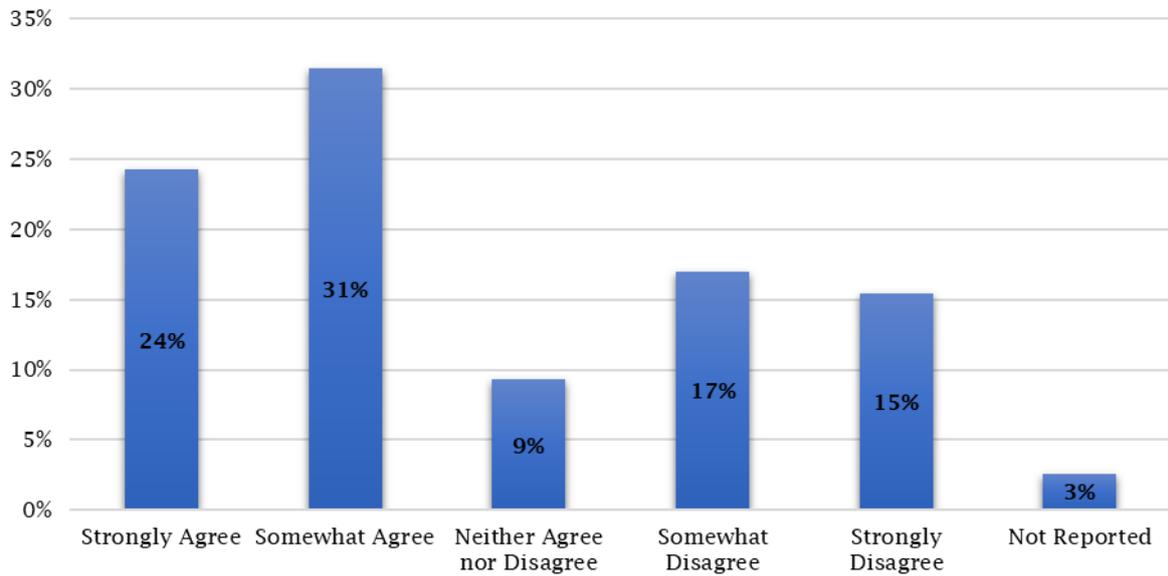

Figure 8: Participants are Familiar with Predictive Tools

Table 13: Country Distribution of Participants

| SN | Country | Respondents | Percent | Cum |
|---|---|---|---|---|
| 1 | United States | 57 | 29.4% | 29.4% |
| 2 | Australia | 23 | 11.9% | 41.2% |
| 3 | Italy | 17 | 8.8% | 50.0% |
| 4 | Canada | 11 | 5.7% | 55.7% |
| 5 | United Kingdom | 11 | 5.7% | 61.3% |
| 6 | Germany | 6 | 3.1% | 64.4% |
| 7 | Netherlands | 5 | 2.6% | 67.0% |
| 8 | Switzerland | 5 | 2.6% | 69.6% |
| 9 | Belgium | 4 | 2.1% | 71.6% |
| 10 | Saudi Arabia | 4 | 2.1% | 73.7% |
| 11 | Spain | 4 | 2.1% | 75.8% |
| 12 | Sweden | 4 | 2.1% | 77.8% |
| 13 | Brazil | 3 | 1.5% | 79.4% |
| 14 | China | 3 | 1.5% | 80.9% |
| 15 | Turkey | 3 | 1.5% | 82.5% |
| 16 | Colombia | 2 | 1.0% | 83.5% |
| 17 | India | 2 | 1.0% | 84.5% |
| 18 | Japan | 2 | 1.0% | 85.6% |
| 19 | Lebanon | 2 | 1.0% | 86.6% |
| 20 | Lithuania | 2 | 1.0% | 87.6% |
| 21 | Palestine | 2 | 1.0% | 88.7% |
| 22 | Poland | 2 | 1.0% | 89.7% |
| 23 | Portugal | 2 | 1.0% | 90.7% |
| 24 | Taiwan | 2 | 1.0% | 91.8% |
| 25 | Europe | 6 | 3.1% | 94.8% |
| 26 | Asia | 4 | 2.1% | 96.9% |
| 27 | Africa | 3 | 1.5% | 98.5% |
| 28 | South America | 3 | 1.5% | 100.0% |
| Total | | 194 | 100% | |

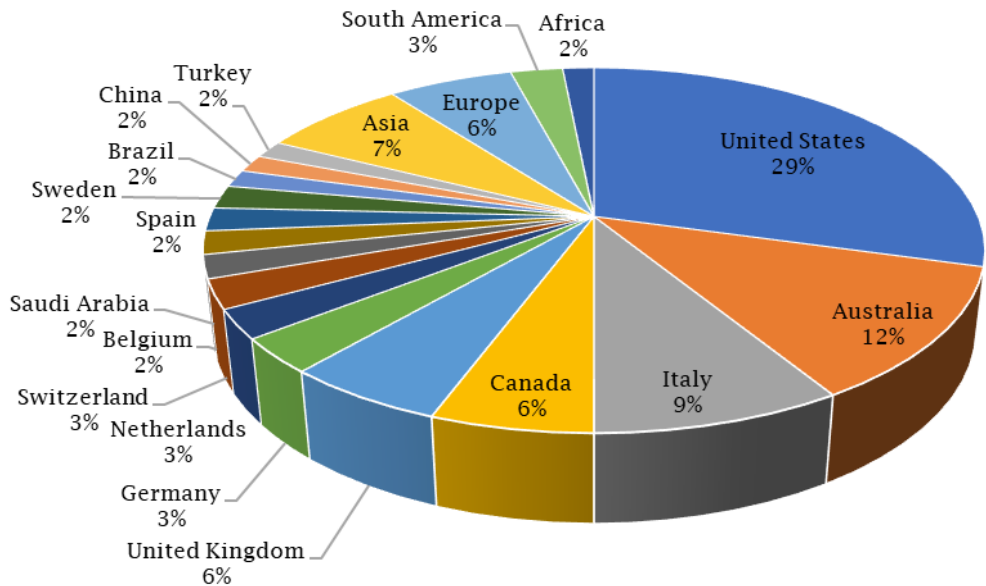

Figure 9: Country Distribution of Participants

## *7.4. Survey Screenshots*

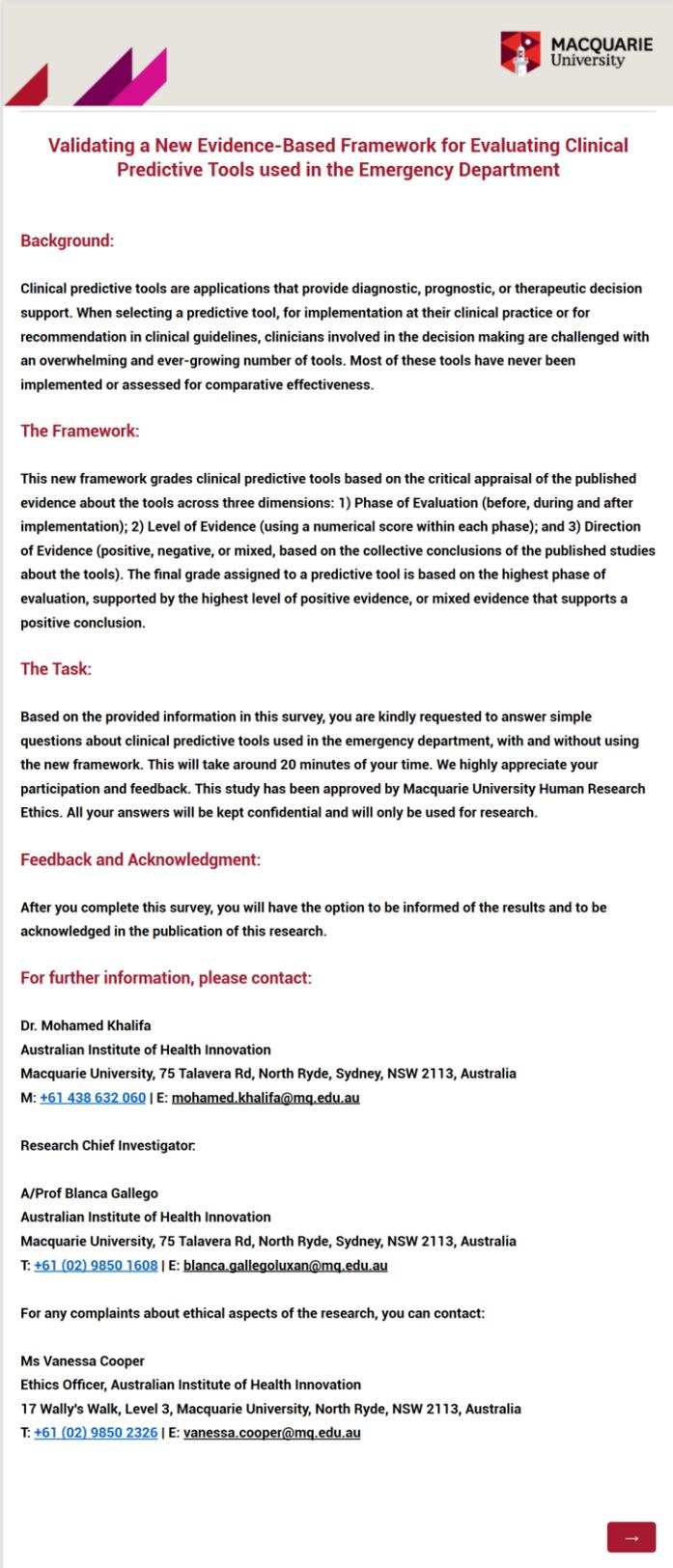

Section 1: The survey introduction

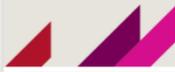

# Paediatric Head Injury Predictive Tools - Evidence-Based Summary

Assume you are the head of a busy emergency department and you are responsible for choosing a predictive tool, for the diagnosis of paediatric head injury, to be implemented at your department or to be recommended in the clinical guidelines. You have a list of five tools and you have an Evidence-Based Summary about the tools:

- PECARN Rule (Paediatric Emergency Care Applied Research Network)
- CHALICE Rule (Children's Head injury ALgorithm for the prediction of Important Clinical Events)
- CATCH Rule (Canadian Assessment of Tomography for Childhood Head injury)
- Palchak (UC Davis) Rule for Paediatric Head Trauma
- Atabaki Rule for Paediatric Head Injury/Trauma

| Tool | Tool Information | | | | | Impact After Implementation | | | Planning for Implementation | | | Performance Before Implementation | | |
|---|---|---|---|---|---|---|---|---|---|---|---|---|---|---|
| | Country | Year | Citations | Studies | Tool Grade | Experimental Studies | Observational Studies | Subjective Studies | Potential Effect & Usability | Potential Effect | Usability | External Validation Multiple Times | External Validation Only Once | Internal Validation |
| | | | | | | A1 | A2 | A3 | B1 | B2 | B3 | C1 | C2 | C3 |
| PECARN | USA | 2009 | 886 | 24 | A2 | | ◐ | | | ● | | ● | | ● |
| CHALICE | UK | 2006 | 308 | 15 | B2 | | | | | ◐ | | ● | | ● |
| CATCH | USA | 2006 | 321 | 12 | C1 | | | | | | | ● | | ● |
| Palchak | USA | 2003 | 247 | 3 | C2 | | | | | | | | ● | ● |
| Atabaki | USA | 2008 | 111 | 1 | C3 | | | | | | | | | ● |

Evidence Direction:
● Positive Evidence
○ Negative Evidence
◐ Mixed Evidence Supporting Positive Conclusion
◑ Mixed Evidence Supporting Negative Conclusion

You can download the full evidence-based report on the five tools from this link: **Full Report**

### Which predictive tool would you choose?

PECARN   CHALICE   CATCH   Palchak   Atabaki   I don't know
  ○         ○        ○        ○         ○          ○

### Regarding my decision in selecting this tool:

| | Strongly agree | Somewhat agree | Neither agree nor disagree | Somewhat disagree | Strongly disagree |
|---|---|---|---|---|---|
| I made this decision based on guessing. | ○ | ○ | ○ | ○ | ○ |
| I made this decision based on my knowledge or experience | ○ | ○ | ○ | ○ | ○ |
| I made this decision based on the information provided in this survey. | ○ | ○ | ○ | ○ | ○ |
| I am confident in my decision. | ○ | ○ | ○ | ○ | ○ |
| I am satisfied with my decision. | ○ | ○ | ○ | ○ | ○ |

Section 2: Block 1: Paediatric Head Injury Predictive Tools With GRASP

## Adult Head Injury Predictive Tools - Briefing Information

Assume you are the head of a busy emergency department and you are responsible for choosing a predictive tool, for the diagnosis of adult head injury, to be implemented at your department or to be recommended in the clinical guidelines. You have a list of five tools and you have briefing information about each tool. You can also download the original study of each tool form the links.

### CCHR – The Canadian CT Head Rule
The tool was developed by Dr. Ian Stiell in Canada in 2001. The tool predicts the need for brain CT imaging in adults after minor head injury/trauma. This diagnostic tool is designed to be used at the emergency department by physicians to decide which patients need CT scan and/or acute intervention. Download the study: CCHR

### Ibañez Model for Head CT
The tool was developed by Dr. Javier Ibañez in Spain in 2004. The tool predicts the need for brain CT imaging in adults after minor head injury/trauma. This diagnostic tool is designed to be used at the emergency department by physicians to decide which patients need CT scan and/or acute intervention. Download the study: Ibañez Model

### KHR – The Kimberley Hospital Rule
The tool was developed by Dr. Abraham Bezuidenhout in South Africa in 2013. The tool predicts the need for brain CT imaging to diagnose trauma-related and non-trauma related acute intracranial pathology conditions in resource-limited environments. This diagnostic tool is designed to be used at the emergency department by physicians to decide which patients need CT scan and/or acute intervention. Download the study: KHR

### Miller Criteria for Head CT
The tool was developed by Dr. Erik Miller in the United States in 1997. The tool predicts the need for brain CT imaging to diagnose minor head trauma in adults. This diagnostic tool is designed to be used at the emergency department by physicians to decide which patients need CT scan and/or acute intervention. Download the study: Miller Criteria

### NOC – New Orleans Criteria
The tool was developed by Dr. Micelle Haydel in the United States in 2000. The tool predicts the need for brain CT imaging after adult head injury/trauma. This diagnostic tool is designed to be used at the emergency department by physicians to decide which patients need CT scan and/or acute intervention. Download the study: NOC

**Which predictive tool would you choose?**

| CCHR | Ibanez | KHR | Miller | NOC | I don't know |
|---|---|---|---|---|---|
| ○ | ○ | ○ | ○ | ○ | ○ |

**Regarding my decision in selecting this tool:**

|  | Strongly agree | Somewhat agree | Neither agree nor disagree | Somewhat disagree | Strongly disagree |
|---|---|---|---|---|---|
| I made this decision based on guessing. | ○ | ○ | ○ | ○ | ○ |
| I made this decision based on my knowledge or experience. | ○ | ○ | ○ | ○ | ○ |
| I made this decision based on the information provided in this survey. | ○ | ○ | ○ | ○ | ○ |
| I am confident in my decision. | ○ | ○ | ○ | ○ | ○ |
| I am satisfied with my decision. | ○ | ○ | ○ | ○ | ○ |

Section 2: Block 2: Adult Head Injury Predictive Tools Without GRASP

## Paediatric Head Injury Predictive Tools - Briefing Information

Assume you are the head of a busy emergency department and you are responsible for choosing a predictive tool, for the diagnosis of paediatric head injury, to be implemented at your department or to be recommended in the clinical guidelines. You have a list of five tools and you have briefing information about each tool. You can also download the original study of each tool form the links.

### Atabaki Rule for Paediatric Head Injury
The tool was developed by Dr. Shireen Atabaki in the United States in 2008. The tool identifies children at low risk of brain injury after minor head trauma. This diagnostic tool is designed to be used at the emergency department by physicians to decide which patients need CT scan and/or acute intervention. Download the study: Atabaki Rule

### CATCH Rule (Canadian Assessment of Tomography for Childhood Head injury)
The tool was developed by Dr. Martin Osmond in the United States in 2010. The tool predicts clinically significant head injury in children after minor head trauma. This diagnostic tool is designed to be used at the emergency department by physicians to decide which patients need CT scan and/or acute intervention. Download the study: CATCH Rule

### CHALICE Rule (Children's Head injury ALgorithm for the prediction of Important Clinical Events)
The tool was developed by Dr. Joel Dunning in the United Kingdom in 2006. The tool predicts death, need for neurosurgical intervention or CT abnormality in children with minor head trauma. This diagnostic tool is designed to be used at the emergency department by physicians to decide which patients need CT scan and/or acute intervention. Download the study: CHALICE Rule

### Palchak (UC Davis) Rule for Paediatric Head Trauma
The tool was developed by Dr. Michael Palchak and Dr. Nathan Kuppermann in the United States in 2003. The tool identifies children at low risk of brain injury after minor head trauma. This diagnostic tool is designed to be used at the emergency department by physicians to decide which patients need CT scan and/or acute intervention. Download the study: Palchak Rule

### PECARN Rule (Paediatric Emergency Care Applied Research Network)
The tool was developed by Dr. Nathan Kuppermann in the United States in 2009. The tool identifies children at very low risk of clinically important brain injury after minor head trauma. This diagnostic tool is designed to be used at the emergency department by physicians to decide which patients need CT scan and/or acute intervention. Download the study: PECARN Rule

**Which predictive tool would you choose?**

| Atabaki | CATCH | CHALICE | Palchak | PECARN | I don't know |
|---|---|---|---|---|---|
| O | O | O | O | O | O |

**Regarding my decision in selecting this tool:**

|  | Strongly agree | Somewhat agree | Neither agree nor disagree | Somewhat disagree | Strongly disagree |
|---|---|---|---|---|---|
| I made this decision based on guessing. | O | O | O | O | O |
| I made this decision based on my knowledge or experience. | O | O | O | O | O |
| I made this decision based on the information provided in this survey. | O | O | O | O | O |
| I am confident in my decision. | O | O | O | O | O |
| I am satisfied with my decision. | O | O | O | O | O |

Section 2: Block 3: Paediatric Head Injury Predictive Tools Without GRASP

Section 2: Block 4: Adult Head Injury Predictive Tools With GRASP

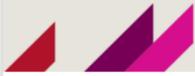

# The Usability of the Evidence-Based Summary

**Regarding the Evidence-Based Summary displayed in the previous screen: How much do you agree with each of the following:**

|  | Strongly agree | Somewhat agree | Neither agree nor disagree | Somewhat disagree | Strongly disagree |
|---|---|---|---|---|---|
| I think I would like to use the Evidence-Based Summary frequently. | O | O | O | O | O |
| I found the Evidence-Based Summary unnecessarily complex. | O | O | O | O | O |
| I think the Evidence-Based Summary was easy to use. | O | O | O | O | O |
| I think I need the support of an expert person to be able to use the Evidence-Based Summary. | O | O | O | O | O |
| I found the functions in the Evidence-Based Summary well integrated. | O | O | O | O | O |
| I think there was too much inconsistency in the Evidence-Based Summary. | O | O | O | O | O |
| I think most people would learn to use the Evidence-Based Summary very quickly. | O | O | O | O | O |
| I found the Evidence-Based Summary very difficult to use. | O | O | O | O | O |
| I felt very confident using the Evidence-Based Summary. | O | O | O | O | O |
| I need to learn a lot of things before I can use the Evidence-Based Summary. | O | O | O | O | O |

**Do you find the evidence-based summary useful? And why?**

Section 3: GRASP System Usability Scale and Usefulness

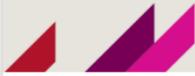

### Demographic Information

**My role in the hospital is best described as:**

- ○ Physician – Consultant
- ○ Physician – Registrar
- ○ Physician – Resident
- ○ Nurse – Senior
- ○ Nurse – Junior
- ○ Other roles

**My specialty is:**

- ○ Emergency medicine / department
- ○ Other specialties / services

**My gender is:**

- ○ Male
- ○ Female
- ○ I prefer not to say

**My age is:**

- ○ Less than 25
- ○ 25 - 34
- ○ 35 - 44
- ○ 45 - 54
- ○ 55 - 64
- ○ 65 - 74
- ○ 75 or older

**My experience in years:**

- ○ Less than 5
- ○ 5 - 9
- ○ 10 - 14
- ○ 15 - 19
- ○ 20 - 24
- ○ 25 - 29
- ○ 30 - 34
- ○ 35 or more

**I am familiar with the head injury clinical predictive tools:**

- ○ Strongly agree
- ○ Somewhat agree
- ○ Neither agree nor disagree
- ○ Somewhat disagree
- ○ Strongly disagree

Section 4: Participants' Demographics

**Feedback and Acknowledgement:**

☐ I would like to receive a feedback on the results of this survey.
☐ I would like to be acknowledged in the publication of this study.

**My name is:**

**My email is:**

All your answers will be kept confidential and will only be used for the purposes of research. Your information is collected only to give you a feedback and will be kept confidential and separate from your answers.

**For further information, please contact:**

Dr. Mohamed Khalifa
Australian Institute of Health Innovation
Macquarie University, 75 Talavera Rd, North Ryde, Sydney, NSW 2113, Australia
M: +61 438 632 060 | E: mohamed.khalifa@mq.edu.au

**Research Chief Investigator:**

A/Prof Blanca Gallego
Australian Institute of Health Innovation
Macquarie University, 75 Talavera Rd, North Ryde, Sydney, NSW 2113, Australia
T: +61 (02) 9850 1608 | E: blanca.gallegoluxan@mq.edu.au

For any complaints about ethical aspects of the research, you can contact:

Ms Vanessa Cooper
Ethics Officer, Australian Institute of Health Innovation
17 Wally's Walk, Level 3, Macquarie University, North Ryde, NSW 2113, Australia
T: +61 (02) 9850 2326 | E: vanessa.cooper@mq.edu.au

Section 5: Participants' Feedback and Acknowledgment